\documentclass[twocolumn,showpacs,preprintnumbers,amsmath,amssymb]
{revtex4-1}
\usepackage{graphicx}
\usepackage{dcolumn}
\usepackage{array}

\begin{document}


\title{Thermally Activated Magnetization and Resistance Decay during Near Ambient Temperature Aging of Co Nanoflakes in a Confining Semi-metallic Environment}

\author{Gregory.~G.~Kenning$^*$, Christopher Heidt, Aaron Barnes, James Martin, Benjamin Grove and Michael Madden.}

\affiliation{Department of Physics, Indiana University of Pennsylvania\\
Indiana, Pennsylvania 15705-1098}

\date{\today}

\begin{abstract}
\setlength{\baselineskip}{20pt} 
We report the observation of magnetic and resistive aging in a self assembled nanoparticle system produced in a multilayer Co/Sb sandwich. The aging decays are characterized by an initial slow decay followed by a more rapid decay in both the magnetization and resistance. The decays are large accounting for almost $70\%$ of the magnetization and almost $40\%$ of the resistance for samples deposited at 35 $^oC$. For samples deposited at 50 $^oC$ the magnetization decay accounts for $\sim 50\%$ of the magnetization and $50\%$ of the resistance.  During the more rapid part of the decay, the concavity of the slope of the decay changes sign and this inflection point can be used to provide a characteristic time. The characteristic time is strongly and systematically temperature dependent, ranging from $\sim1$x$10^2~s$ at 400K to $\sim3$x$10^5~s$ at 320K in samples deposited at $35~^oC$. Samples deposited at 50 $^oC$ displayed a 7-8 fold increase in the characteristic time (compared to the $35~^oC$ samples) for a given aging temperature, indicating that this timescale may be tunable. Both the temperature scale and time scales are in potentially useful regimes. Pre-Aging, Scanning Tunneling Microscopy (STM) reveals that the Co forms in nanoscale flakes. During aging the nanoflakes melt and migrate into each other in an anisotropic fashion forming elongated Co nanowires. This aging behavior occurs within a confined environment of the enveloping Sb layers. The relationship between the characteristic time and aging temperature fits an Arrhenius law indicating activated dynamics.

\end{abstract}

\maketitle


\section{Introduction}
\label{sec:SclTemp1}
\setlength{\baselineskip}{20pt} 
Since the early 1980s the ability to produce nanostructured magnetic materials has lead to a variety of new and useful effects. The discovery of  Giant Magnetoresistance (GMR) in Fe/Cr multilayer sandwiches in the late 1980s ushered in a new era in electronics with a solid-state device which on a local scale can effectively perform the same function as the Faraday effect \cite{Grunberg86, Carbone87}.  Producing functionality in the solid-state can offer many advantages including miniaturization and cost.  The formation of nanoparticles and nanoparticle assemblies offer an added dimension to the mix as nanoparticles often show significantly different properties than their bulk counterparts. Berkowitz 
et al.\cite{Berkowitz92} produced heterogeneous Cu-Co alloys which as deposited showed spinglass like behavior. In these types of materials the competing long range RKKY ferro and antiferromagnetic exchange interactions coupled with the spatial disorder of the Co, in the matrix, produce a spin glass state for 
$<25\%$ Co \cite{Childress91}. They also found that after enhanced aging (annealing at high temperature), phase separation occurs producing Co rich particles with an average diameter of 4 nm and an average spacing of 8 nm. The system then displayed GMR. There have been many other studies on noninteracting and interacting Co nanoparticles monodispersed on surfaces and in matrices, particularly insulating matrices\cite{Huttel04, Binns05}. Most of these studies have been on spherical nanoparticles which have optimized structural stability (minimized surface to volume ratio). Kleemann et al.\cite{Kleemann01} have reported the observation of superferromagnetism in discontinuous 2D arrays of CoFe nanoparticles embedded 
in $Al_2O_3$.  The 1.3 nm/3 nm $CoFe/Al_2O_3$ layer thickness used in that manuscript is similar to the 1.5 nm/2.5 nm Co/Sb used in this study. The major difference of course is that $Al_2O_3$ is insulating and Sb is semi-metallic.  

Aging is the process through which materials, not in thermodynamic equilibrium, evolve on their way to attaining thermodynamic equilibrium. Systems in thermodynamic equilibrium have, on average, static physical parameters and hence do not age.  The aging process can generally be divided\cite{Henkel09} into two types: 1) Physical Aging, where temporal effects occur in a material phase and can be reset by changing thermodynamic parameters to reset the phase.  Over the last few decades, the study of physical aging in materials has developed with many materials showing interesting time dependent changes, often over specific and tunable timescales. Since Struik$'$s work\cite{Struik78} on aging in glassy polymers, time dependent studies have been performed on a wide range of  materials displaying physical aging including spin glasses\cite{Chamberlin83} and electronic glasses\cite{Ovadyahu03} among others.  2) Chemical Aging is often associated with an irreversible change of material properties due to chemical or thermal evolution. Chemical aging at ambient temperatures, for example in alloys, is often much too slow to be experimentally quantified\cite{Marder87}.  In some cases where an understanding of time dependence can have important economic or cultural value, for example in pharmaceuticals\cite{Waterman05}, food storage\cite{Cruz09},  document preservation\cite{Begin02} and many other examples\cite{Nelson90} a technique called accelerated aging is often employed to bring aging effects over long times (usually months or years) down to laboratory time scales. Accelerated aging requires subjecting the samples to extremes in temperature, light, humidity etc. in order to speed up the aging process.  In materials processing, thermally accelerated aging is  called annealing and has been used as a metellurgical technique for millenia. Annealing promotes atomic migration, crystallite growth and the elimination of defects. 

In this paper we report aging experiments which follow the time dependent magnetization and resistivity of a Co/Sb multilayer deposited with an e-beam evaporation technique. We further investigate the time dependence as a function of measuring temperature. STM and magnetization measurements  are used to analyze the structural and magnetic states of the system before during and after the aging process. While the technique of evolving materials as a function of temperature is clearly annealing, the observation of this process is more akin to the observation of aging phenomena. Throughout this manuscript we use the term aging to describe the temporal evolution of our samples at a given temperature.

\section{Experimental Methods}
\label{sec:SclTemp2}
The samples used in the study were deposited in a vacuum of $\leq1\times10^{-6}$ Torr in an Edwards model E306A e-beam evaporator.   The deposition rate was monitored insitu during the entire course of the deposition with a deposition monitor. Deposition rates were approximately .2-.3 nm/minute. The vast majority of the samples discussed in this paper were produced through deposition on the $<111>$ face of silicon although we also report data for a few samples deposited on $Al_2O_3$ and common commercial aluminum foil. Since the Edwards comes with a single e-beam gun, multilayer samples were produced by evaporating the Sb and then rotating the Sb 99.999$\%$ target out of the gun and the Co 99.9$\%$ into the gun to produce a layer of Co.  Our goal in this study was to produce Co nanoparticles in the solid state within a semimetallic environment. Co is known to form magnetic nanoparticles in a variety of environments and has a bulk resistivity of $62.7 n\Omega \bullet m$ at 300 K\cite{HBCP}. Bulk Co has a melting temperature of 1,768.15 K. Sb is a semimetal and has a bulk resistivity of $417 n\Omega \bullet m$, almost seven times higher than that of Co. Bulk Sb has a melting temperature of 903.78 K  We waited approximately 5 min. after the Co was deposited in order to maximize time for possible Volmer-Weber particle growth mechanisms to act.  All samples reported in this paper were made with nominal layering of Co/Sb 1.5 nm/ 2.5nm.  Aging decays in samples with 10 layers, 50 layers and 100 layers were analyzed and all show reasonably similar effects. STM measurements were made on samples with 1 and 1.5 layers. Unless specifically mentioned, all data represent measurements on separate samples.

Special care is required for the handling and storage of materials that show aging properties at room temperature. Samples with 10 layers required about two hours to make. As the number of multilayers increased to 50-100 layers, depositions could extend over several days. While we believe that aging begins during the deposition process, aging at room temperature was quite slow. Data on samples presented in this paper were deposited on substrates held at $35~^oC$ and $50~^oC$. Since measuring temperatures and deposition temperatures overlap, to avoid confusion we report deposition temperatures in $^oC$ and measuring temperatures in K.  The characteristic time was estimated to be approximately 3 weeks for samples deposited at $35~^oC$ and approximately 6 months for samples deposited 
at $50~^oC$. Care was however taken to minimize any type of aging that may have occured before the sample was actually measured. Once removed from the evaporation chamber, the samples were placed in an argon glovebox where they were removed from the evaporation sample holder and separately placed in specimen holders in an argon environment. The samples were then stored at 77 K. When needed, a particular sample was removed, warmed to room temperature in a dry box and either placed directly into the apparatus for magnetoresistance measurements or attached to the sample holder with GE varnish for magnetization measurements. The varnish was allowed to dry for four hours before placing the sample in the LakeShore Model 7307 Vibrating Sample Magnetometer (VSM) with Magnetoresistance (MR) option. 

The sample insertion and recording techniques for the magnetization and resistance measurements were slightly different and will be discussed. For the magnetization measurements the cryostat was preheated to the measuring temperature and the magnetic field was set to 0 T. The sample and sample rod were placed in the cryostat, attached to the head drive(which vibrates the sample), the head drive  turned on and an initial zero field magnetization measurement recorded. Here there were some sample to sample deviations. For example, individual zero field magnetization measurements (50 layers) varied from -132 $\mu$emu  to 300 $\mu$emu although most samples had zero-field magnetizations of a few $\mu$emu to a few tens of $\mu$emu. The largest measured values correspond to  approximately 10-15$\%$ of the maximum magnetization ($\approx2.2-2.5~m$emu) obtained once the field is turned on. A .01 T magnetic field was then turned on, setting the time $t=0$ s for the aging decay. Once the field was set, a large rapid rise in the magnetization was observed that maximized in a few tens of seconds. The signal then began to decay. During the magnetization decay the magnetization was measured every 10 s. The measurements reported here were made with both the magnetic field and sensing coils directed into the plane of the multilayers. In a magnetic field of .01 T, the out-of-plane magnetic measurements show similar time dependent results but with a maximum magnetization of about 25$\%$ of the in-plane magnetization. The magnetic moments therefore appear to be directed mainly in-plane. Bulk Co is hexagonal and the easy direction is the c axis\cite{Morrish65}.

For the majority of resistance measurements the sample and MR probe were inserted into a preheated cryostat, set at the measuring temperature, and into a preset magnetic field of .01T. Several resistance vs. time measurements were also made in zero magnetic field. Once the program was initiated it took approximately 130s for the apparatus to take the first point. The measurements reported here are made with the Lakeshore MR Option which is a 4-probe method using four in-line tantalum leads to pierce through the multilayers. This geometry leads to the current flowing in-plane. The magnetic field was also directed in-plane. Two measurements (current parallel to, and 180$^o$ to the magnetic field) were made and averaged for each point.

After observing the aging decays we attempted to reinitialize the samples to determine if we were dealing with chemical or physical aging. Holding the sample (35 $^oC$ deposited) at 400 K in 1 T for 24 hours did not  reset the sample back to its initial condition.  We also tried placing the samples (for short periods of time 10 s-60 s) into a preheated oven over a range of temperatures varying from 700 K to 1100 K. We were not able to reset the sample and began to conclude that chemical aging was likely.

We set out to eliminate the possibility that chemical aging occurs due to the substrate or due to oxygen annealing.  Although solid state dynamics are arguably too slow (in this temperature range 300-400 K over measuring our timescales $\sim1$x$10^5~s$)) for Si in the substrate or oxygen (samples were open to air) to diffuse through the 10-100 multilayers, it was considered. We should also add that either of these scenarios would likely increase the resistance not decrease it. Samples were produced on Si, $Al_2O_3$ and Al foil with similar and consistent decay effects observed  on each (see Figure 3). We believe that this rules out the substrate as a major mechanism for the decay. Oxygen could also affect the magnetic moments if it diffused into these thin film samples. Oxygen could enable antiferromagnetic coupling of the Co moments through the superexchange interaction\cite{Morrish65}.  While we have not been able to rule this scenario out with magnetic experiments we did place a 100nm Au cap on a sample (as a barrier to O penetration) before aging it and found effectively the same aging behavior.  This rules out Oxygen annealing during measurement as the mechanism for the decay. The other possibility is that O combines with the Co during deposition in the vacuum chamber although the level of the vacuum suggests that this effect should not be significant.

We initially set out to produce nanoparticles of Co in a solid state Sb matrix with growth formation through a Volmer-Weber growth mechanism. In an earlier study, we deposited samples at room temperature and gave significant time (5 min) after Co deposition to maximize nanoparticle growth. Every 5 multilayers, the system was cooled for 30 minutes to maintain a deposition temperature of approximately room temperature. We found in these samples magnetic signatures of 6-8nm Co nanoparticles (Peak in ZFC, ZFC-FC remanence, 100K-200K blocking temperature\cite{Jun08}, strong field dependence of peak, no waiting time effect).  While these results will be discussed elsewhere, it was clear that we were producing small Co particles and not perfect layers.  Low angle x-ray data show no evidence of superlattice peaks corroborating the lack of ideal layering. We should also point out that after aging, the nanoparticle signature disappeared. 

\section{Results}
\label{sec:SclTemp3}

\begin{figure} 
\centering 
\resizebox{\columnwidth}{2.5in}{\includegraphics{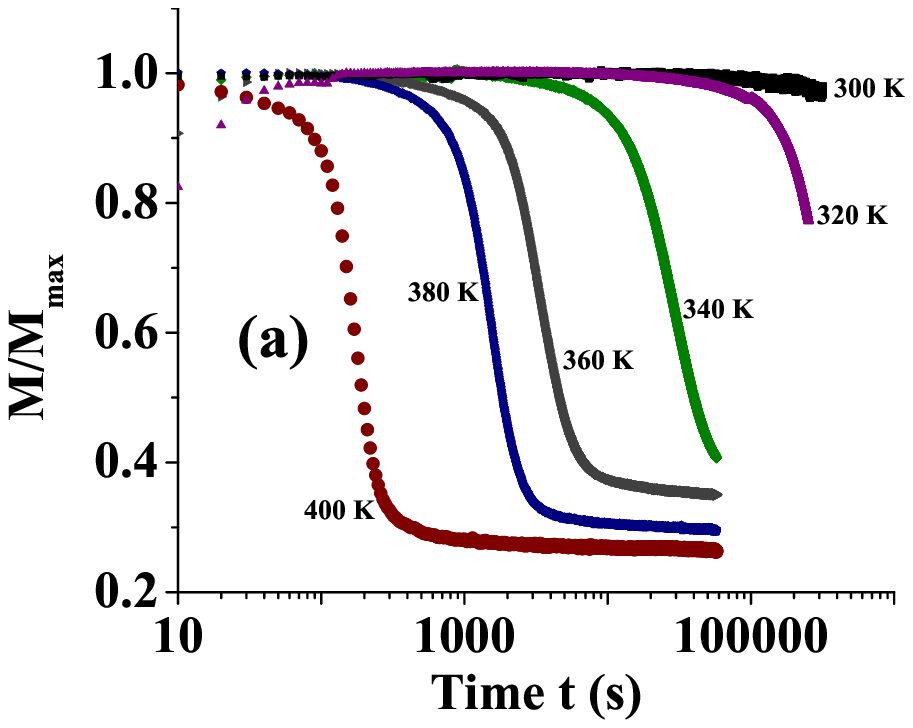}}
\resizebox{\columnwidth}{2.5in}{\includegraphics{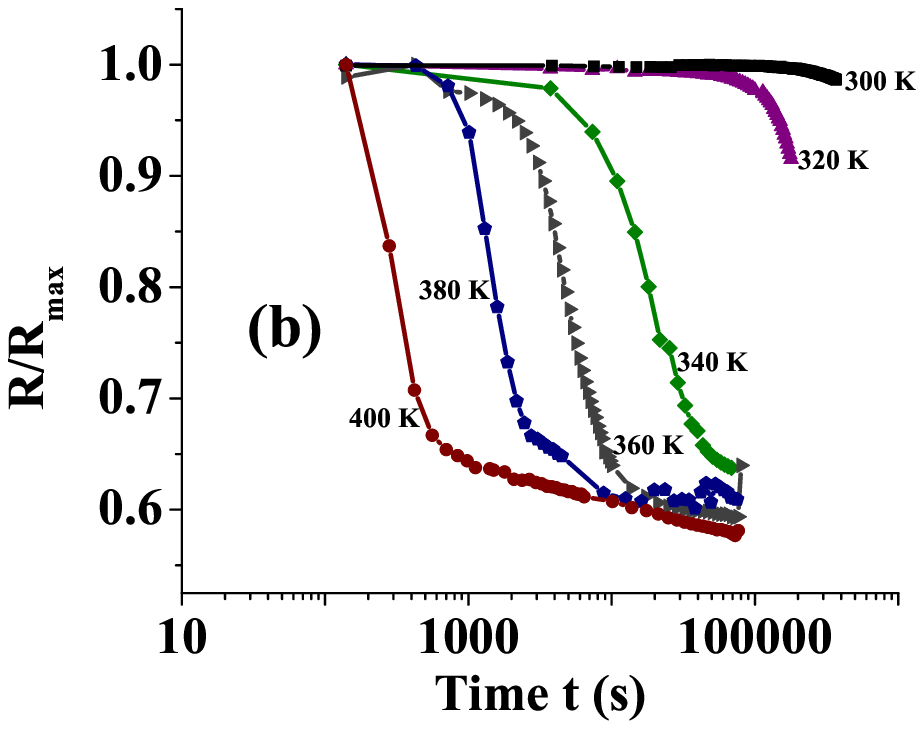}}

\caption{\label{fig:bob1} Magnetization and Resistance Aging of Co/Sb, a) Normalized magnetization decay curves for a set of temperatures ranging from 300 K to 400 K. The curves were normalized to their maximum value. Samples were deposited at 
35 $^oC$ on $<111>$ face of Si.  b) Normalized resistance decay curves (measured in a magnetic field of .01 T) for a set of temperatures ranging from 300 K to 400 K. The curves were normalized to their maximum value. Samples were deposited at 35 $^oC$ on $<111>$ face of Si.}
\end{figure}

Figures 1a) and 1b) display aging decay curves for the magnetization and resistance, as a function of temperature, for samples deposited at 35 $^oC$. Following the work of Struick\cite{Struik78} we plot the data on a logarithmic time scale. We characterize the observed aging decays with two timescales.  The first time scale represents birth or the beginning of aging.  In the Co/Sb samples we believe that the process of aging begins very shortly after production.  At room temperature the decay is still quite slow, allowing for transport and storage of the samples (at 77K) with little effect on the measured aging curves. For these experiments we set the birth timescale ($t$=0s) to the time when the sample was placed in the preheated cryostat and the measuring program initiated. We define the second time scale as the time at which the concavity of the slope of the decay changes sign (i.e. the inflection point of the decay curves). This time scale has previously been employed by Nordblad et al.\cite{Nordblad87} and we will utilize it to determine an effective age for each decay curve. For both the magnetization and resistance this time will be called 
$\tau_i$. Figure 1 shows that both the magnetization and resistance of the Co/Sb system undergo  decays which shifts significantly as a function of temperature. For samples deposited at 35 $^oC$ the magnetization curves undergo a total decay of approximately 50-70$\%$ of the initial magnetization. Every increase of 20K brings an $\sim$5-10 fold increase in $\tau_i$. The resistance decays display a similar time dependence and show a decrease of approximately 40-50$\%$ of the initial resistance.

\begin{figure} [t] 
\centering
\resizebox{3.0in}{7.0in}{\includegraphics{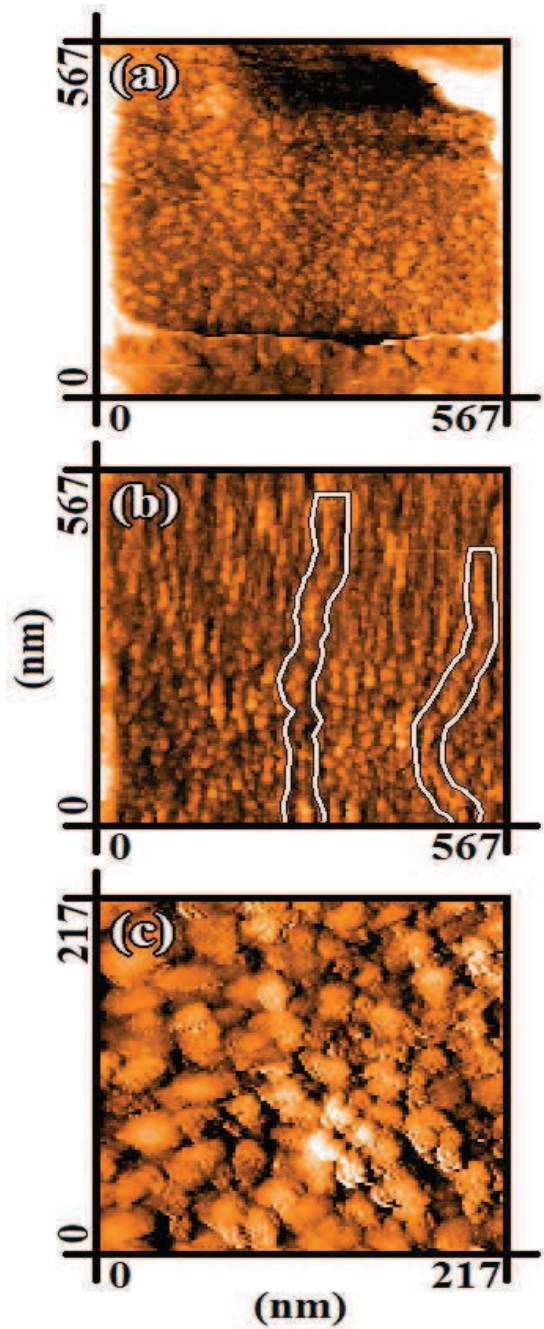}} 
\caption{\label{fig:bob4} STM Images, a) STM image of pre-aged Sb/Co/Sb (2.5 nm/1.5 nm/2.5 nm).    b) STM image of same sample as in a), after aging.  c) STM image of Sb/Co ((2.5nm/1.5nm). The Co is on the surface. }
\end{figure}

Scanning Tunneling Microscopy (STM) was performed on a variety of samples, including samples before and after aging. From Figure 2a) and 2b) the results of aging are directly observable. From Figure 2a), it appears that the Sb wets the surface of the substrate and provides a uniform covering. It also appears that we can observe the Co particles through the upper layer. This sample shows both a microcrack near the bottom of the image and film edge on the right hand side. (We included this image with these features to give perspective; none of the other samples or regions of samples we imaged displayed cracks.) Statistical analysis  of randomly chosen areas of the preaged samples was used to quantify the Co particles. Approximately 120 particles were analyzed. Since the particles are not perfectly spherical, the maximum length across each particle was chosen for analysis. While maximum lengths ranged from a few nanometers to almost 40 nm the distribution was approximately Gaussian with a mean maximum length of 19 nm and a standard deviation of  5 nm. The  nearest neighbor(nn) distance distibution  was limited at short distance simply by the resolution of the technique. Our analysis however gives a mean nn distance of 4.0 nm with a standard deviation of 1.7 nm. Profile analysis indicates that the Co particles are on average 2.5 nm in height. Therefore the lateral dimensions are much larger than the thickness and the particles are effectively nanoflakes or nanopancakes. Image analysis of different regions shows that the Co attains approximately 50-70$\%$ coverage (variation over different areas analyzed). These 
nanoflakes are self-assembling and have a large surface to volume ratio. This ratio is likely responsible for most of the effects observed in this paper. Figure 2b) is the same sample (different area) as Figure 2a) after being heated to 400 K for 20 min. This is approximately the time required for a sample  deposited at 50 $^oC$ to fully age i.e. full decay of large decay portion of the magnetization and resistance (see Figure 4). It can be observed that there is a drastic difference in the structure of the material. It appears that the Co has crept anisotropically, through the Sb, forming long nanowires. Statistical analysis of these wires determined a mean width of 15 nm and lengths ranging from from 100 nm to $\geq$567 nm. It should be pointed out that smaller nanoparticles observed in the same image were not included in this analysis and the maximum length measured was limited by the image size. The mean length determined was 330 nm with a standard deviation of 130 nm.  Profile scans suggest that the nanowire thicknesses are on the order of 2.3 nm. It also appears that there is significant overlap of these nanowires both along their widths and along their lengths and that they may be in contact. At this stage it is unclear what produces the directionality of the wires. 

Figure 2c) displays an Sb/Co single bilayer system with the Co as the top layer deposited at $50~^oC$. The sample is unaged in the sense that once removed from the deposition chamber, it was stored overnight at 77 K and then measured at room temperature $\sim$2 hours after being removed from the dewar. Statistical analysis indicate that the nanoparticles are approximately 30nm in size and 1.6nm in thickness producing 91$\%$ coverage. The nanoflakes appear to overlap and seem to have aged isotropically even though no heat treatment was applied. We therefore conclude that the enveloping Sb layer confine the nanoparticles, slowing down migration and producing the time effects evident in Figure 1.

\section{Discussion}
\label{sec:SclTemp4}

 The time and temperature dependence of the decays in these samples indicate that the process may be thermally activated.  In nature when a system in one state can transition to another state by  thermally activated hopping over an energy barrier, the system can generally be described by an Arrhenius law.  As written for chemical dynamics, the Arrhenius equation determines a frequency of transition k, between states at temperature $T$. 

\begin{eqnarray}
k =A exp(-E_a/k_BT)
\end{eqnarray}

The time dependence for the state transition is given by the Neel-Arrhenius form\cite{Neel49}, that was originally used to describe the characteristic spin flip time (barrier hopping time) of superparamagnetic nanoparticles. 
\begin{eqnarray}
\tau=\tau_o exp(E_a/k_BT),
\end{eqnarray}

In Eq. 2), $E_a$ is the activation energy or barrier height, $\tau_o$ is the fluctuation timescale and $\tau$ is the characteristic hopping time at temperature $T$. In the case of superparamagnetic particles, $E_a= KV$ where K is the anisotropy constant and V the particle volume although in general this formulation gives the characteristic hopping timescale for any two-state system  separated by a single energy barrier. In an ensemble of two state systems $\tau$ provides an averaged characteristic time scale.

While the nanoparticles in these samples likely have activated dynamics governing spin flipping, a more likely scenario, considering the change in morphology, is softening/melting of the nanoflakes and thermally activated migration (creep). The large surface to volume ratio of these flakes should strongly enhance melting. The nucleation and growth of crystallites during aging has been extensively studied since the 1920s (for a review see Rios et al.\cite{Rios05}). It is clear from the STM images that one and likely two growth mechanisms are present. First, it is clear that portions of the nanoflakes are migrating through the intervening Sb to couple to each other. The model for the velocity of the interface of high angle boundaries is thermally activated\cite{Rios05}. Driving and retarding forces to this mechanism have also been considered. These forces may be relevant for the creep of the Co through the Sb. Second, once the Co particles form an interface with each other, the process of coalescing can reduce the free energy by including a rotation of the crystal axis of one or both of the particles to form a single crystal. While both of these mechanisms have been applied to annealing of polycrystalline samples it is likely that they can also be applied to the Co/Sb system.

\begin{figure} [t] 
\centering 
\resizebox{\columnwidth}{3.0in}{\includegraphics{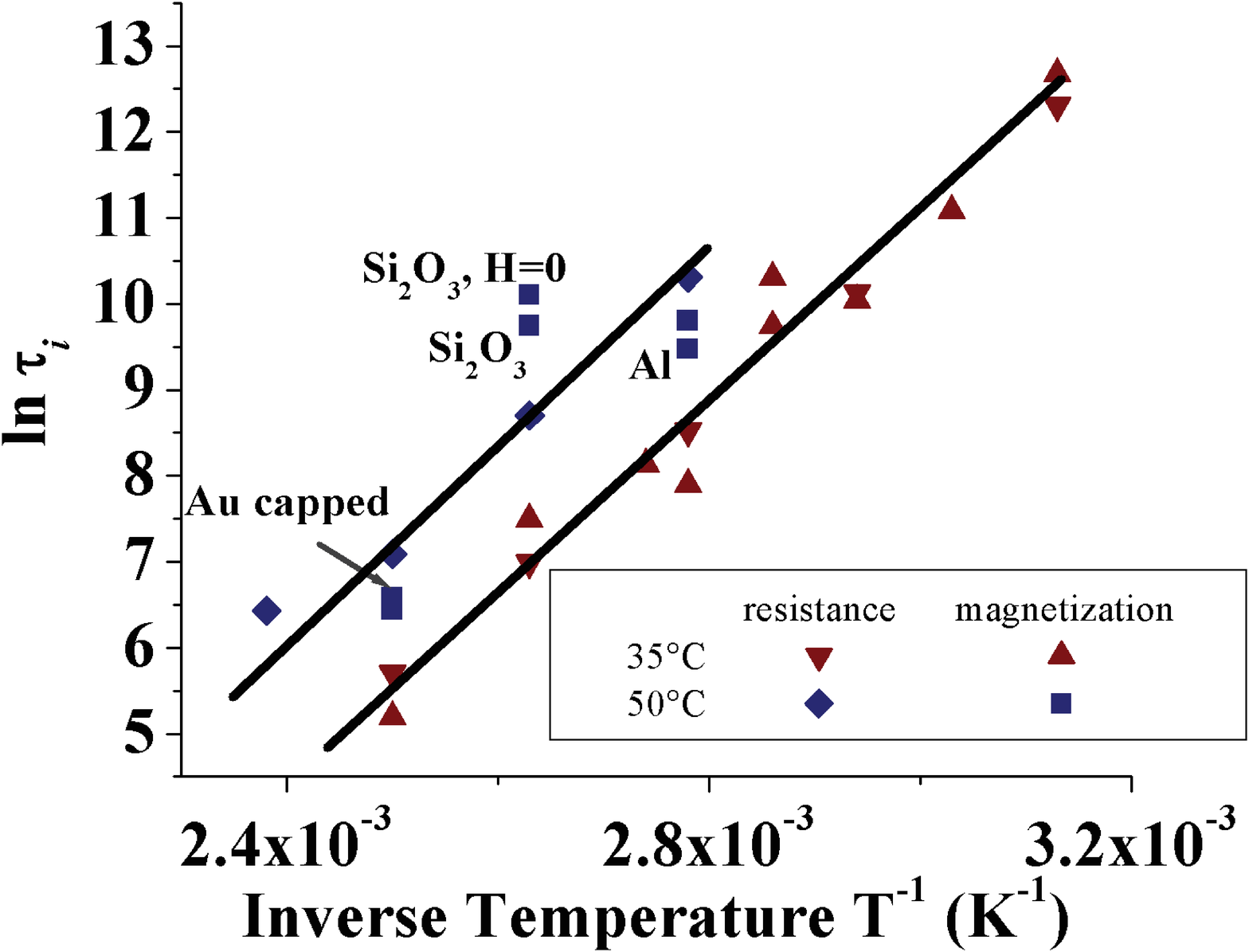}}
\caption{\label{fig:bob3} Activated Dynamics Analysis of Aging, a) Ln $\tau_i$ vs. 1/T for all samples decays.  Unless indicated, all samples were deposited on the $<111>$ face of Si and measured in a field of .01 T.  Red symbols indicate samples deposited at 35 $^oC$. Blue symbols indicate samples deposited at 50 $^oC$. }
\end{figure}

In Figures 3)  we plot ln$\tau_i$ vs. 1/T  for two sets of samples deposited at $35~^0C$ and 
$50~^0C$.  $\tau_i$ is determined from the inflection point of the decay curves. Samples at both deposition temperatures include sets of $\tau_i$ values from both magnetization and magnetoresistance decays. It can be observed that the decays from both the magnetization and resistance  have compatible time scales. Fluctuations in $\tau_i$ data provide a better metric of the reproducibility of time scales than say the error in the determination of $\tau_i$, which are generally small. It would appear that a range of characteristic time scales may be achieved, at a single temperature, by depositing the samples at different temperatures. Annealing at the interface between subsequent Sb layers (i.e. between Co particles) may explain the deposition temperature dependence.  Extrapolating,
 ${1\over T} \rightarrow 0$, gives the value of $\tau_o$, the thermal fluctuation timescale.  The value of $\tau_o$ obtained for fits over samples from the two deposition temperatures is approximately 1-4 x 10$^{-10}$s.  From the graph of sample deposited at 35 $^oC$ we find an activation energy of .91 eV.

An understanding of the decay in the resistance must consider the following results:
1) The measured resistivity of the unaged sample is approximately one order of magnitude greater than the expected resistance (calculated neglecting interfacial mixing and/or scattering) of this particular multilayer sandwich. 
2) Magnetoresistance measurements on an unaged samples at 300K  show less than 1$\%$ change from $ 1~T \rightarrow -1~T \rightarrow 1~T$. Therefore even though at first glance this system (magnetic/metallic multilayer) may appear to be a GMR system, it is not. 
3) Even though the  majority of the resistance decay measurements reported here were made in .01 T, to remain consistent with the magnetization measurements, we find that we get similar decays in zero magnetic field (see Figure 3, upside down triangles). Therefore, this is not a magnetoresistive effect but a resistive effect.

As mentioned previously, the measured resistance of the unaged sample is approximately one order of magnitude greater than what we would expect from a calculation of the expected resistance of this particular multilayer sandwich. In Fig. 2a) we observe a microcrack near the bottom of the sample. A sample with large numbers of dislocations or a fractured sample could produce a large resistance but as previously mentioned, with the exception of Fig. 2a), we do not observe any cracks and surface profiling suggests that the Sb forms a contiguous layer so it is unlikely that dislocations are the reason for this enhanced resistance. Two effects may contribute to raise the resistance of the Co nanoflakes to the flow of conduction electrons.  First, in a normal metal, current is carried equally by electrons of both spin components. In a ferromagnet, current is carried preferentially by the majority spin component. van Son et al.\cite{Son87} have shown that at a ferromagnetic/non-ferromagnetic metal interface, this difference produces an electrochemical mismatch that increases the boundary resistance. Current flowing through a Co nanoparticle would have to pass two interfaces which would act like an additional pair of series resistors, significantly increasing the resistance of each nanoparticle.   Second, spin orbit scattering from the Sb/Co interface is also likely to significantly increase the scattering cross-section, increasing the resistivity.  Upon the advent of aging, the nanoparticles begin to flow into each other. An electron needs to cross only one interface to enter the extended low resistivity Co nanowires where it can be transported with little resistance for hundreds of nanometers. This is likely the short-circuit which decreases the resistance.

\begin{figure} [t] 
\centering 
\resizebox{\columnwidth}{4.4in}{\includegraphics{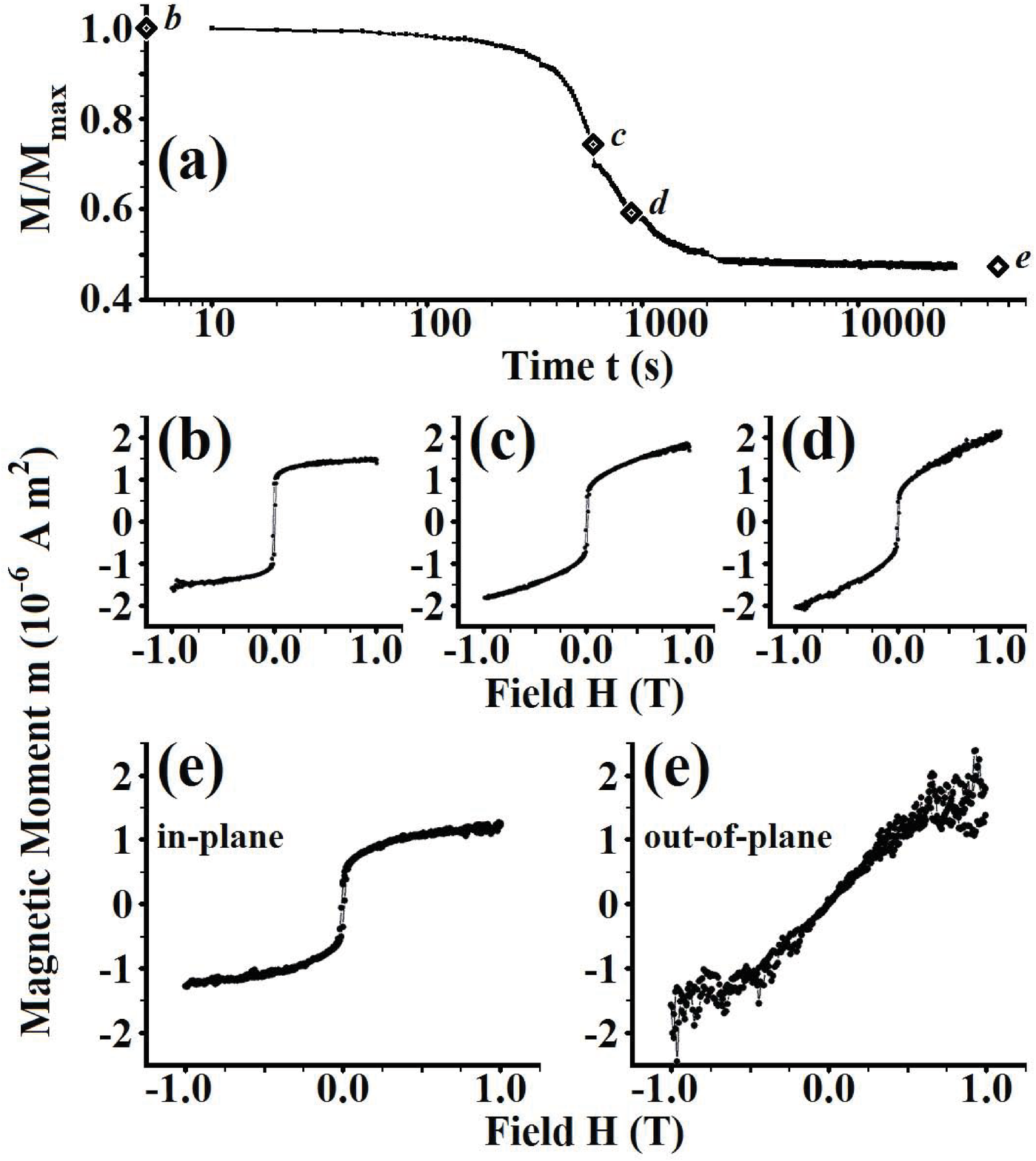}}

\caption{\label{fig:bob2} Hysterisis Measurements During the Aging Process, All measurements on a single sample. Samples were deposited at 50 $^o$C on the $<111>$ face of Si. Aging was performed at 400 K. a) Assembled magnetization decay measured in a magnetic field of .01 T. b) c) d)and e) indicate the points where the corresponding hysterisis points were taken at 300 K. }

\end{figure}

Experimental analysis of the hysteresis loops  taken at various times during the aging decay provides insight on possible mechanisms for the magnetization decays. In Fig. 4, we display results (on a single sample) for a systematic study of the magnetic hysteresis during the decay process. The sample analyzed was deposited at 50 $^oC$. The sample was aged at 400 K for set periods of time and the magnetization decay measured.  After aging, the sample was removed from the cryostat to room temperature. The cryostat was then cooled to 300 K and the sample placed back in the cryostat for the hysteresis measurements at 300 K. During hysteresis measurements, the sample would often be held at 300 K for up to 12 hours. At 300 K the estimated characteristic time (extrapolation of Figure 3) is on the order of six months, so we assumed aging would be much slower and that we could ignore the effects due to aging during the hysteresis measurements. This retardation in time scale is evident in Fig. 2a where the segments making up the set of decays measured at 400 K are reconstructed into a single decay.  

We see that the hysteresis loops of the unaged samples have a ferromagnetic signature with a coercive field of approximately 0.006 T. The unaged sample also has a remanent magnetization of $\sim$1.1 $memu$ and a slightly increasing slope in the high field regime. At 1 T the magnetization reaches a maximum of approximately 1.5 $memu$. After aging at 400 K for 600 s, we observe changes in the hysteresis loops (Fig. 4c). While the coercive field remains similar to the unaged sample there is a distinct tilt to the hysteresis loop, the remanent magnetization decreases to .8 $memu$ and the slope in the high field magnetization increases. These changes are indicative of a possible paramagnetic or frustrated contribution. Both of these effects are accentuated with further aging. Figures 4c) and 4d) are taken during the rapid decrease in the magnetization.  After aging well past the rapid decay and well into a much slower, possibly logarithmic decay, the in-plane hysteresis again shows ferromagnetism with a reduced saturation moment. Interestingly, the out-of-plane magnetization, which displayed a ferromagnetic behavior before aging, in this sample, has a (somewhat noisy) paramagnetic signature after aging. Considering the shape of the particles we cannot rule out significant demagnetization effects in the out-of-plane data.
There are many mechanisms that can cause a decrease in the magnetization. The hysteresis measurements as well as the STM data provide clues to the magnetization decrease during aging. First, the unaged samples look ferromagnetic.  While superferromagnetism is a possibility, magnetization vs. temperature studies down to 10 K show no evidence of ultra small paramagnetic particles forming between the nanoparticles during the deposition, as observed in CoFe/$Al_2O_3$ \cite{Bedanta07}. As a matter of fact the saturation magnetization can be fit to the standard $T^{3/2}$ form between 10 K and 300 K. The STM results on unaged samples indicate nanoparticles that are extended in-plane ($\sim 25-30 nm$) while the out of plane thickness is only $\sim 2-3 nm$.  This geometry significantly decreases the in-plane demagnetizing field, stabilizing the particle against domain wall formation. Therefore, we believe that it is likely that, before aging, the nanoflakes behave as single domain ferromagnetic particles. As the samples age these flakes begin to flow and contact  each other. As the nanowires begin to form it is likely that the enhanced sizes will increase the demagnetizing field, inducing anti-parallel allignment of domains and the formation of domain walls. This is likely the mechanism which reduces the magnetization during aging. In the initial stages of nanoflake contact two factors must be considered. First, in the  initial contact the energy of a domain wall would be minimized if it formed at the much smaller contact interface as opposed to in the interior of a particle. Second, the c-axis of the individual nanoflakes are likely misaligned in the initial contact\cite{Rios05} which may lead to frustration between the moments. This frustration may be the cause of the tilting of the hysteresis loops during aging.

Finally a word on sample to sample fluctuations in the time dependent data. It may be possible to construct this system accurately enough to use the time dependent decay as a solid state clock and a switch. However, at this time the fluctuations in $\tau_i$ are too large to make this practical. From the size data of the STM images we estimate approximately $10^{11}$ nanoflakes in a ten layer sample of $20~mm^2$. This number should provide significant statistical averaging for production of a more accurate and reproducible characteristic time scale. The e-beam evaporator used in this study effectively produces a point source of evaporating atoms and our substrates were laid out in a plane intersecting the sphere of that source. Sample to sample deviations are therefore to be somewhat expected. Variations in nanoflake sizes and separation distances also likely contribute to inhomogeneities in the time scale. There are a variety of techniques available for controlling sample sizes\cite{Majetich06} and separations\cite{Jun08} in spherical nanoparticles which may be useful but it is unclear whether they may be useful for nanoflakes.

In conclusion, we report a multilayer system which shows distinct aging properties as a function of temperature. Aging is observed as a large change in both the magnetization and resistance as a function of time. In particular a specific time can be associated with each decay and hence the system acts like a clock and a switch that, at a particular time (dependent on the temperature), can be observed to switch off or on. The timescales over which this particular time occurs in our samples range from seconds to months.  The temperatures, at which these timescales are observed, range from just above room temperature up to  120$^o$C. 

The authors would like to thank E. Zaremba, P. Sibani, D. Talwar, K. Komjati and R. Orbach for useful discussions.


\begin{thebibliography}{99}

\setlength{\baselineskip}{20pt} 
\bibitem{Grunberg86}	P. Gr\"{u}nberg, R. Schreiber, Y. Pang, M. B. Brodsky, and H. Sowers  Physical Review Letters 57 (19): 2442–2445, (1986). 

\bibitem{Carbone87} C. Carbone and S. F. Alvarado   Physical Review B 36 (4: 2433(1987),
M. N. Baibich , J. M. Broto, A. Fert, F. Nguyen Van Dau, F. Petroff, P. Eitenne, G. Creuzet, A. Friederich, and J. Chazelas  Physical Review Letters, , v61, p2472. (1988)
%
\bibitem{Berkowitz92} A. E. Berkowitz, J. R. Mitchell, M. J. Carey, A. P. Young, S. Zhang, F. E. Spada, F. T. Parker, A. Hutten, and G. Thomas , Phys. Rev. Lett. 68, 3745–3748 (1992)
%

\bibitem{Childress91}J. R. Childress and C. L. Chien,  Phys. Rev. B 43, 8089–8093 (1991) 

%
\bibitem{Huttel04} Y. Huttel, H. Gómez, C. Clavero, A. Cebollada, G. Armelles, E. Navarro, M. Ciria, L. Benito, J. I. Arnaudas, and A. J. Kellock, J. Appl. Phys. 96, 1666 (2004)  
%
\bibitem{Binns05} C. Binns, K N Trohidou, J Bansmann, S H Baker, J A Blackman, J-P Bucher, D Kechrakos, A Kleibert, S Louch, K-H Meiwes-Broer, G M Pastor, A Perez and Y Xie, J. Phys. D: Appl. Phys. 38 R357 (2005)
%
\bibitem{Kleemann01} W. Kleemann, O. Petracic, Ch. Binek, G. N. Kakazei, Yu. G. Pogorelov, J. B. Sousa, S. Cardoso and P. P. Freitas, Phys. Rev. B 63, 134423 (2001) 
%


\bibitem{Henkel09} Malte Henkel, Haye Hinrichsen,  Sven Lübeck,  Non-Equilibrium Phase Transitions, Volume 1, Series: Theoretical and Mathematical Physics, Jointly published with Canopus Publishing Limited, Bristol, UK, 2009, IV,  

%
\bibitem{Struik78} Struik, L. C. E., 1978 Physical aging in amorphous polymers and other materials / L. C. E. Struik Elsevier Scientific Pub. Co.
%
\bibitem{Chamberlin83}	R.V.Chamberlin, M.Hardiman and R.Orbach, J. Appl. Phys., 5, 1771 (1983). L.Lundgren, P.Svedlindh, P.Nordblad and O.Beckman, Phys. Rev. Lett., 51, 911(1983); L.Lundgren, P.Svedlindh, P.Nordblad and O.Beckman, J. Appl. Phys. 57, 3371 (1985), G. F. Rodriguez, Kenning, G.G. and R. Orbach,  Physical Review Letters, Vol. 91, No. 3, 037203-1, July 18,  2003. 
%
\bibitem{Ovadyahu03}	Z. Ovadyahu, Slow Relaxations and nonequilibrium dynamics in condensed matter Les Houches, 2003, Volume 77/2003, 213-231, DOI: 10.1007/978-3-540-44835-8-11 Course 11: Non-equilibrium Dynamics and Aging in the Electron Glass 

%
\bibitem{Marder87} M. Marder, Physical Review A 36, 858–874 (1987) 

%
\bibitem{Waterman05}
Waterman KC, Adami RC.
Int J Pharm. 2005 Apr 11;293(1-2):101-25.
%
\bibitem{Cruz09}
S. A. Cruz, M. Zanin, M. A. B. de Moraes
Journal of Applied Polymer Science
Volume 111, Issue 1, pages 281–290, 5 January 2009
%
\bibitem{Begin02}
P L Begin, E Kaminska 
Restaurator International Journal For The Preservation Of Library And Archival Material (2002)
Volume: 23, Issue: 2, Pages: 89-105
%
\bibitem{Nelson90}
W Nelson, 
Accelerated Testing: Statistical Models, Test Plans, and Data Analysis, (Wiley Series in Probability and Mathematical Statistics) 
 Wiley, 1990. 



%
\bibitem{HBCP}Handbook of Chemistry and Physics, 56th Edition, Robert C. Weast (Ed.), CRC Press (1975) 


%
\bibitem{Morrish65}A. H. Morrish, The Physical Principles of Magnetism, Wiley: New York,. 1965
%
\bibitem{Jun08} Young-wook Jun, Jung-wook Seo and Jinwoo Cheon, Accounts of Chemical Research, 2008, 41 (2), pp 179–189
%

\bibitem{Nordblad87} P. Nordblad, P. Svedlindh, P. Granberg, and L. Lundgren, Phys. Rev. B 35, 7150–7152 (1987) 
%
\bibitem{Neel49}L. Neel,  Ann. Geophys., vol. 5, pp. 99-136, 1949. 
%
\bibitem{Rios05} Paulo Rangel Rios, Fulvio Siciliano Jr, Hugo Ricardo Zschommler Sandim, Ronald Lesley PlautI, Angelo Fernando Padilha, Mat. Res. vol.8 no.3 São Carlos, REVIEW ARTICLE 2005
%
\bibitem{Son87}P. C. van Son, H. van Kempen, and P. Wyder, 
Phys. Rev. Lett. 58, 2271–2273 (1987) 

%
\bibitem{Bedanta07}S. Bedanta, T. Eimüller, W. Kleemann, J. Rhensius, F. Stromberg, E. Amaladass, S . Cardoso, and P. P. Freitas, Physical Review Letters 98, 176601 (2007)
%
\bibitem{Majetich06} Majetich, S, and Sachan, M, J. Phys. D, 39, R407-R422 (2006)
Magnetostatic Interactions in Magnetic Nanoparticle Assemblies: Energy, Time, and Length Scales, 


\end{thebibliography}
\end{document}